\documentclass{JINST}
\usepackage{amsmath}

\title{Interest Point Detection for Reconstruction in High Granularity Tracking Detectors}

\author{B. Morgan \\
Department of Physics, \\
University of Warwick, \\
Coventry, CV4 7AL, United Kingdom\\
E-mail: \email{Ben.Morgan@warwick.ac.uk}}

\abstract{%%
This paper presents an investigation of the use of interest point detection 
algorithms from image processing applied to reconstruction of interactions in 
high granularity tracking detectors. Their purpose is to extract keypoints from
the data as input to higher level reconstruction algorithms, replacing the role
of human operators in event selection and reconstruction guidance.
Simulations of $\nu_{\mu} + ^{40}Ar \rightarrow \mu^{-} + p$ in a small liquid
argon time projection chamber are used as a concrete example of a modern high
granularity tracking detector. Data from the simulations are used to 
characterize the localization of 
interest points to physical features and the efficiency of finding interest 
points associated with the primary vertex and track ends is measured. 
A high degree of localization is found, with $93\%$ of detected interest 
points found within 5mm of a physical feature. Working in two 2D projections, 
the primary vertex and both track ends are found in both projections in 
$85\%$ of events. It is also shown that delta electrons can be detected.

}


\begin{document}
%%
%% Body text for tracking calorimeter HSP paper
%%

\section{Introduction}
Bubble chambers are the classic example of a high granularity tracking detector,
and provided a key tool for many discoveries in particle
physics~\cite{kalmus:bcreview,giacomelli:bcreview}.
The complex event topologies studied by these experiments and the limited
computing power of the time presented a great challenge in reconstructing and
characterizing the recorded interactions~\cite{villemoes:bc,mermikides:bc}.
Significant progress was made in fully automated measurement for very simple 
event topologies~\cite{polly:bcscanner}, yet more complex interactions still 
required a high degree of human input for pattern recognition and measurement
guidance~\cite{numu:hbc}. In these cases, an operator would
examine the image on a scanning table, identifying events and marking
keypoints such as vertices to provide guidance to the automated
measurement system~\cite{gouache:erasme}. 

As electronic detectors took over from the bubble chambers, the general trend 
in tracking moved to trackers layered around a known vertex region 
with trajectories measured at a few sparse points with very high resolution.
Pattern recognition techniques have correspondingly evolved
to handle this data and complete automation has been achieved~\cite{mankel:recon:review}. Yet
significant interest has remained in high granularity tracking due the potential
for robust particle identification and precise kinematics in a 
simple, homogeneous detector. The modern exemplar of such a detector is the
Liquid Argon Time Projection Chamber (LAr-TPC)~\cite{Rubbia:117852} which
provides simultaneous tracking and calorimetry with millimetric granularity.
Some reconstruction and pattern recognition techniques have been demonstrated
for LAr-TPCs, but like bubble chambers these have used human-computer 
interaction to both select events and to guide the
reconstruction~\cite{martinez:recon:thesis, ankowski:icarus:em}. With the start
of data-taking for
ICARUS T600~\cite{menegolli:icarus2010} and ArgoNeut~\cite{antonello:uslar},
significant interest has developed in fully automating reconstruction for 
LAr-TPCs, both to remove operator bias and to handle the large datasets expected.

The use of human-computer techniques in LAr-TPC reconstruction has arisen from 
many of the same challenges faced by bubble chamber analyses. 
LAr-TPCs are targeted at neutrino oscillation and cross-section studies at the 
few GeV scale~\cite{menegolli:icarus2010,antonello:uslar,cavanna:xsec}, and also
nucleon decay searches~\cite{lartpc:pdecay}. The event classes to be reconstructed thus 
consist of multiple vertex hadron decays and electromagnetic/hadronic showers,
giving a complex pattern recognition task. Like a bubble chamber, interactions 
can occur anywhere within the TPC volume, giving no clue to primary and
secondary vertex locations and consequently complicating initialization of any 
pattern recognition algorithm. Unlike the separated
tracking and calorimeter components of typical collider detectors, tracks and
showers occur together in a LAr-TPC, leading to a complex ``track'' recognition
task that must separate and cluster both tracks and showers. 
These tasks of vertex/keypoint identification and 
shower/track separation have not yet been automated in LAr-TPC reconstruction 
and are therefore a key goal for any fully computational approach.

Like bubble chamber photographs, LAr-TPC data represents an image of the pattern of
ionization left by charged particles, with reconstruction efforts concentrating
on the use of two 2D (pixel) images projected from the underlying 3D (voxel)
image. Pattern recognition in images is a well established subfield of Image
Processing, to which bubble chamber analyses contributed through the 
Hough Transform for identifying straight line features~\cite{hough}. Many other types
of features can be detected in images, and perhaps the most well known class of
feature detection algorithms are the interest point
detectors~\cite{tuytelaars:review}. An interest point may be defined as a point
in an image which differs in a mathematically describable way from its local
neighbourhood. Whilst many mathematical descriptors have been developed, the
most common and frequently used are those based on local intensity differences.
This concept was first introduced by Moravec~\cite{Moravec::original}, who
defined ``points of interest'' as points in an image where the intensity 
varied strongly in all directions. Interest point detection based on Moravec's
concept is also known as ``corner detection'' as corner like structures display
the strong directional intensity variance picked out by the algorithms. As
structures like decay vertices in a LAr-TPC image essentially form a corner,
this motivates an application of interest point detection to identify the
position of vertices and other keypoints for use by downstream reconstruction
algorithms. This approach does not seek to fit a vertex or keypoint location,
rather, it aims to provide a rough location to guide higher level pattern
recognition, replacing the role of a human operator.

In this paper, intensity variation-based interest point detection is applied to
identify keypoints in $\nu_{\mu} + ^{40}Ar \rightarrow \mu^{-} + p$ events
simulated in a small LAr-TPC. Section~\ref{theory} describes the
quantification of intensity variation in terms of the structure tensor and shows
how this is used to define an interest point response function. Section~\ref{simulation}
describes the simulation of the detector, input events and output data.
Analysis of this data to extract interest points in two 2D images is discussed
in Section~\ref{analysis}, with results on the localization of interest points
with physical features and the efficiency of identifying known keypoints. A
summary of these results and future directions is given in
Section~\ref{conclusions}.

\section{Interest Point Detection using the Structure Tensor}
\label{theory}
Interest point detection using intensity variations in an image was 
first introduced by
Moravec~\cite{Moravec::original}, who defined ``points of interest'' as points 
where the intensity $I(x,y)$ varies strongly in all directions.
Moravec's technique quantifies the local intensity variation $M(x,y)$ around 
each pixel 
by finding the minimum difference in intensity between a small window centred 
on the pixel and the same window shifted by a few pixels along each of the
eight cardinal plus ordinal directions.
Large values of $M(x,y)$ thus correspond to points with large intensity
variations in all directions, with interest points hence identified and
extracted as the coordinates of local maxima of $M(x,y)$.

Harris \& Stephens~\cite{HSP::original} removed the directional quantization inherent in
Moravec's algorithm by expanding the intensity variation equation using a first
order Taylor series. They showed that the intensity gradient structure local to
a point is encapsulated by the structure tensor (or second
moment matrix) averaged over the neighbourhood of the point:

\begin{equation}
S(x,y) = g(\sigma_s)\ast\left[
\begin{array}{cc}
I_x(x,y)^2 & I_x(x,y)I_y(x,y) \\
I_x(x,y)I_y(x,y) & I_y(x,y)^2
\end{array}
\right]
\label{structure:tensor}
\end{equation}

\noindent
where $\ast$ denotes convolution, $I_{x(y)}$ is the partial derivative of the 
image with respect to $x$($y$) and $g$ is a Gaussian window

\begin{equation}
g(\sigma) = \exp \left( \frac{-(x^2+y^2)}{2\sigma^2}\right).
\end{equation}

\noindent
which is convolved with the second moment matrix to provide averaging and reduce
noise~\cite{HSP::original}. Whilst Harris \& Stephens used a Prewitt filter to
calculate the derivatives, modern implementations recommend a Gaussian
derivative filter to improve rotational invariance ($S$ is fundamentally
rotationally invariant, but certain derivative operators are not) and
signal-to-noise~\cite{schmid:hsp}. Use of a Gaussian derivative filter provides
differentiation of the image as

\begin{equation}
I_x(x,y) = g(\sigma_d)\ast \frac{\partial I(x,y)}{\partial x} =  \frac{\partial g(\sigma_d)}{\partial x} \ast I(x,y).
\label{image:derivative}
\end{equation}

\noindent
The standard deviations $\sigma_s$ for averaging and $\sigma_d$ for
differentiation are usually chosen to be the same, though this is not required.

Equation~\ref{structure:tensor} provides the basis for many interest point
detectors (see for instance~\cite{tuytelaars:review}) because its eigenvalues and eigenvectors describe the intensity
variation local to any point in the image. If we consider a line, i.e. a
particle track-like object, then the gradient in intensity will be small along
the line and large perpendicular to the line. Consequently, $S(x,y)$ for
$(x,y)$ along the line will have one large eigenvalue and one small. In
contrast, if we consider two lines meeting in a `V', i.e. a decay vertex-like
structure, then the intensity gradient will be large in all directions at points
close to the `V' vertex. Thus at these points, $S(x,y)$ will have two large
eigenvalues. Functions of $S(x,y)$ can therefore be constructed to produce a
feature response image $R(x,y)=f(S(x,y))$ whose local maxima will correspond to
the interest points of a particular type.

The best known, and most widely used, feature response function is that proposed
by Harris \& Stephens (sometimes known as the Plessey
operator)~\cite{HSP::original}, with

\begin{equation}
R_{HS}(x,y) = \det(S(x,y)) -k\mathrm{Tr}^2(S(x,y))
\end{equation}

\noindent
where $k$ is a tunable parameter. Since
the determinant(trace) corresponds to the product(sum) of the eigenvalues of
$S$, $R_{HS}$ will be large if both eigenvalues are large. Thus maxima of
$R_{HS}$ correspond to points with a corner-like, i.e. vertex-like, structure. 

However, the extra degree of freedom in $R_{HS}$ from the $k$ parameter
complicates the characterization of interest point detection for LAr-TPC reconstruction. To simplify this analysis, another common feature response function 
proposed by F\"orstner~\cite{forstner:detector} and Noble~\cite{noble:phd} is used:

\begin{equation}
R_N(x,y) = \left\{
\begin{array}{l l}
\frac{\det (S(x,y))}{\mathrm{Tr}(S(x,y))} & \quad \text{if $\mathrm{Tr}(S)>0$}\\
0 & \quad \text{if $\mathrm{Tr}(S) = 0$}
\end{array}
\right.
\label{noble:response}
\end{equation}

\noindent
Like the Harris \& Stephens response function, $R_N$ will have maxima at
corner-like structures in the image, and hence interest points are extracted as
the coordinates of these maxima. 

Interest points for reconstruction in a LAr-TPC will clearly be features such as
decay vertices, interaction points and track stop points. As all of these can be
built from primitive line and `V' shapes, processing a LAr-TPC image with
Equation~\ref{noble:response} should lead to these points being picked out as
maxima in the resulting response image. It should be noted that interest
point extraction is not applied here as a fit to the locations of, e.g., 
vertices, as interest points will only be localized by the extraction to the 
nearest pixel. Whilst techniques exist for sub-pixel resolution, the
smoothing and gradients used in structure tensor based approaches can displace
maxima in the response from the true feature location~\cite{tuytelaars:review}.
Rather, interest point detection in LAr-TPC reconstruction should be viewed 
as replacing the role of a human operator in picking out, e.g., rough vertex
locations and other key points for use by higher level pattern
recognition algorithms.

\section{Simulation of Detector, Events and Data}
\label{simulation}
The Geant4 toolkit~\cite{geant4} was used to model a simple LAr-TPC as a
stainless steel cylinder of 1m radius and 2m height, filled with natural argon in
the liquid state. All material properties were
taken from the NIST database built in to Geant4. Standard electromagnetic and
decay processes were modelled for all particles. To provide a simple and well
defined final state topology for characterizing the interest point detector, no
hadronic decay or inelastic processes were modelled. The application of interest point
detection for hadronic decay chains and showers will be the subject of a
future publication.

The GENIE~\cite{genie} software package was used to generate a 1000 event sample
of quasi-elastic charged current (QEL-CC) $\nu_{\mu} + ^{40}Ar \rightarrow \mu^{-} + p$
events. The input $\nu_{\mu}$ spectrum was monoenergetic with an energy of 
0.7GeV, chosen to match that around the peak flux of the JPARC
beam~\cite{besnier:t2krev}. The neutrinos were directed in a beam along the 
x-axis through the centre of the TPC vessel. 
At the chosen energy, events with a pure $\mu^{-}+p$ final state comprise 
around 60\% of all GENIE-generated QEL-CC events.

The liquid argon volume was divided into $1\times1\times1$mm$^3$ voxels, giving
a slightly higher granularity than current detectors. This allows a conservative
study, as higher granularities are more challenging due to the greater detail
present. All primary and secondary particles were tracked through the voxels down to zero
energy or until they left the TPC volume. Energy deposits by charged particles 
passing through the voxels were tallied into a map between voxel coordinates
$(i,j,k)$ and the total energy deposited in that voxel, $E$. Together with this
voxel map of the energy deposits, the positions of physical feature points
were recorded, specifically the primary vertex, stop/decay/exit points of
the primary muon and proton, and the creation and stop/exit points of delta 
electrons. No modelling of the readout system or hit/voxel reconstruction from
raw data was performed as this is highly experiment-specific. 

%% Use test_tpcimage if plots of events and interest points needed

\section{Detection and Analysis of Interest Points from Simulated Events}
\label{analysis}
\begin{figure}
\includegraphics[width=\textwidth]{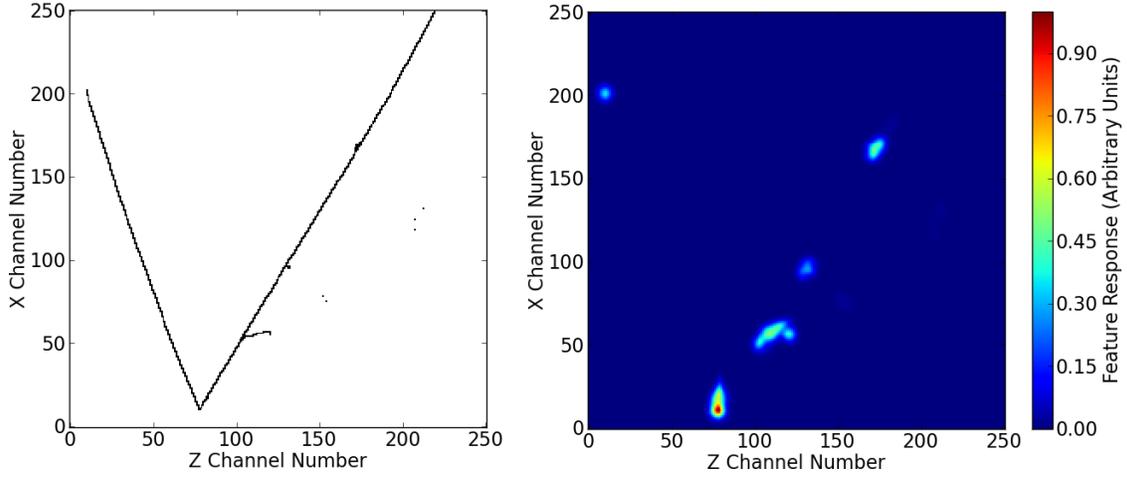}
\caption{XZ projection images of hits (\textit{left}) from a $\nu_{\mu} + ^{40}Ar
\rightarrow \mu^{-} + p$ event ($\mu^-$ right, $p$ left), and the resultant
response image $R_N$ for the hits(\textit{right}). Channel numbers give the pixel
coordinates, with 1 channel equivalent to a spatial dimension of 1mm. The
primary vertex, proton stopping point and delta electrons are clearly picked out
as intensity peaks in the response image.}
\label{event}
\end{figure}

In line with other LAr-TPC reconstruction efforts, the analysis of interest
points uses two 2D projections of the 3D voxel data.
For each event, the 3D energy deposition map was projected to two 2D images with
$1\times1$mm$^2$ pixels in the $xz$ and $xy$ planes perpendicular to the
neutrino beam.
Each non-zero pixel was set to a value of 1 to remove energy
deposition information. Whilst this reduces the total information content in the
image, it specifically highlights the geometric structure of the event which is
of primary importance in locating vertex and other physical interest points.
Leaving the energy information in results in a biased response to regions of
high $dE/dx$ contrast, masking the more important geometric information.
The feature response images $R_N(x,y)$ and $R_N(x,z)$ were
calculated for each input image using
Equations~\ref{structure:tensor},~\ref{image:derivative} and
~\ref{noble:response}. Derivative and smoothing filter widths were chosen
as $\sigma_d = \sigma_s = 2.5$mm. Figure~\ref{event} shows an example event and
the resultant response image.

Interest points were identified as local maxima of the $R_N$ images, with the
pixel coordinates and $R_N$ value of these maxima extracted using a non-maximum suppression filter. This filter used a square window of width 10mm, chosen to 
match twice the total smoothing/derivative width, $\sigma_d+\sigma_s$, the 
rough scale of peaks in $R_N$ for isolated structures. The resultant sets of
interest points, $(x,y,R_N)$ and $(x,z,R_N)$, for each projected image were
finally thresholded to select only those points with a value of $R_N$ within
$90\%$ of the maximum $R_N$ value in the set.

\subsection{Localization of Interest Points}
\label{analysis:distances}
\begin{figure}
\includegraphics[width=.5\textwidth]{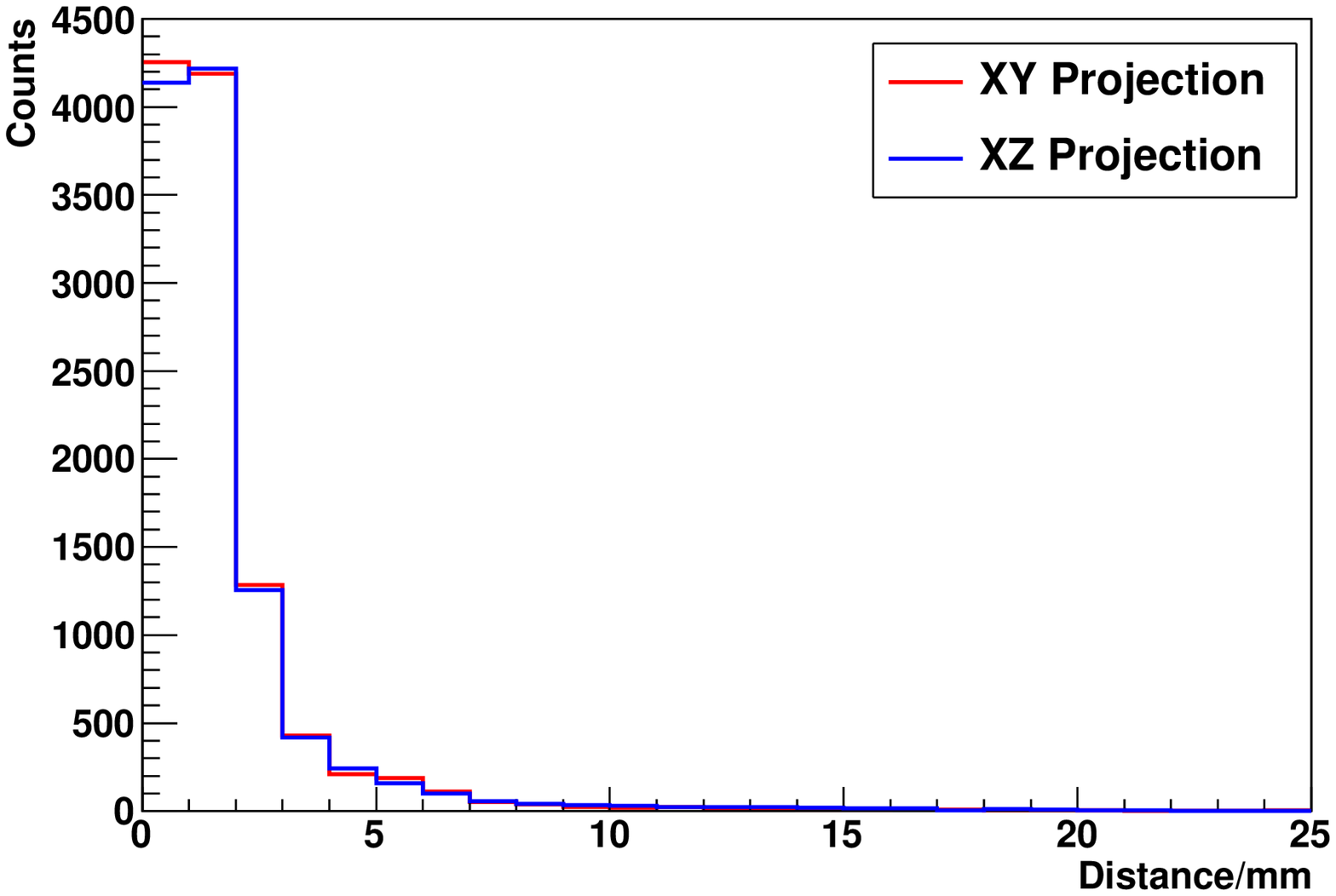}
\includegraphics[width=.5\textwidth]{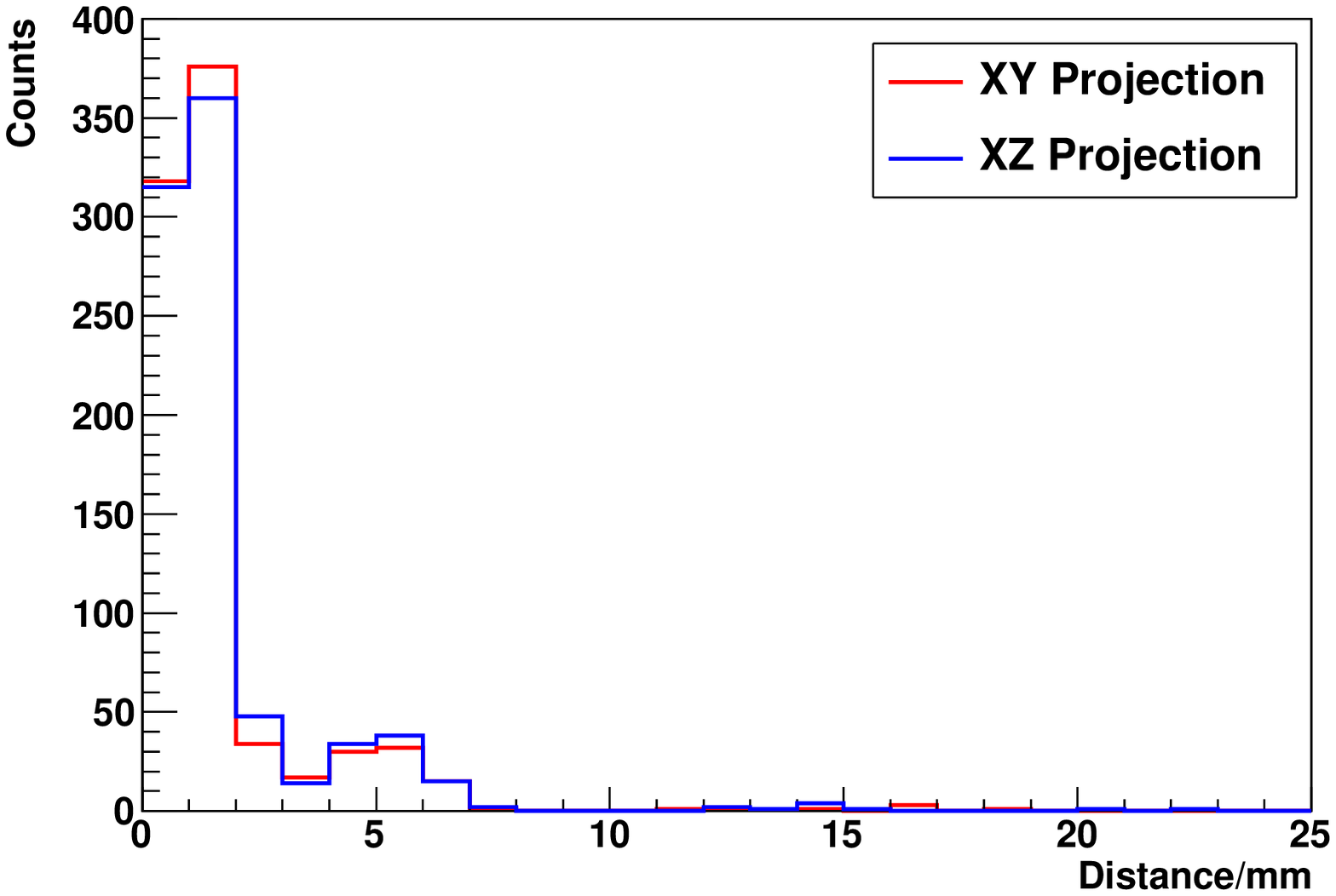}
\caption{Distributions of distances between all detected interest points and 
nearest physical feature (\textit{left}), and interest points nearest to the primary
vertex and the primary vertex itself (\textit{right}), for the $xy$ and $xz$ projections.}
\label{fig1}
\end{figure}

If interest points extracted from the data are to be useful in higher level pattern recognition algorithms, they should be closely related in space to actual 
physical features. To
characterize this, the nearest (projected) physical point to each extracted interest point
was identified from the simulation outputs, with the distance between the points being measured together
with the classification of the physical point. The distribution of these
distances in each projection is shown in Figure~\ref{fig1} for all interest points and for the
subset of interest points whose nearest physical point was the primary vertex.

It can be seen from Figure~\ref{fig1} that the distance distributions are
peaked at small distances, indicating that the detected interest points are correlated with physical features. Both distributions are the
same, within statistical errors, for both projections which is expected as both
planes contain the beamline. The localization of physical features by detected
interest points is good, 93\% of all interest points being within 5mm of a physical feature,
and 89\% of the interest points closest to the primary vertex being within 5mm of the
vertex. Clearly, the detected interest points have high correlation with 
physical features and are thus of potential use as seeds or keypoints by other
pattern recognition algorithms. Algorithmic and physical effects contribute to
the broadening of the distance distributions, though it is again emphasized that
the extracted feature points do not represent a fit to a physical location, e.g.
a primary vertex.

Due to the use of gradients in structure tensor approaches, maxima in the
response function may be displaced from the physical
feature, particularly at `T' junctions~\cite{smith:dphil}. This effect only
contributes to the broadening below $\sim$5mm as this is the typical width of
peaks in the response function, as seen in Figure~\ref{event}. 
At larger distances, ``noise'' features contribute to the broadening due to 
the relatively low response threshold chosen. Tuning of the derivative/smoothing
filter widths and threshold parameters may help to reduce these effects, and this is left for concrete 
implementations as it will be sensitive to the event topology and the behaviour
of other pattern recognition algorithms making use of the detected points.
The shoulder at $\sim5$mm in the distance distribution for interest points closest to 
the primary vertex arises from interference between the gradients of the
primary vertex and nearby delta electrons. This leads to a merging of the
resultant peaks in the response function and consequently a displacement of the
measured interest point location(s) from the primary vertex and delta electron start/end. This
will occur on scales of the overall width of the derivative/smoothing kernels used, leading
to maxima in $R_N$ , and hence identified interest points, displaced on scales
of $\sim$5mm.

\subsection{Efficiency of Detecting the Primary Vertex and Track End Points}
\label{analysis:efficiency}
Whilst Section~\ref{analysis:distances} shows physical features to be located
accurately, the detection algorithm must also find them at high efficiency if they are to be useful to higher level pattern recognition algorithms. 
In particular, the primary vertex and primary track end
points may be key for certain track finding algorithms. To characterize the
efficiency of finding an interest point related to the primary vertex, the number of
events with an interest point classified as closest to the primary vertex (see
Section~\ref{analysis:distances}) was counted for each projection. This number
divided by the total number of events gives the efficiency of directly finding
an interest point related to the primary vertex, and is given in Table~\ref{table1} for
each projection.

\begin{table}
\begin{center}
\begin{tabular}{|l|c|c|}
\hline
Measure & $xy$ Efficiency & $xz$ Efficiency\\
\hline\hline
Vertex Found Directly & $(81.8\pm2.9)$\% & $(82.0\pm2.9)$\% \\
Feature $<5$mm from Vertex & $(90.5\pm3.1)$\% & $(91.1\pm3.1)$\% \\
\hline
Feature $<5$mm from Both Primary Track Ends& $(90.3\pm3.1)$\% & $(88.8\pm3.0)$\% \\
\hline
\end{tabular}
\end{center}
\caption{Efficiencies for finding primary vertex and muon/proton track ends in
the two image projections. Uncertainties were calculated as the Poisson error on
the counted number of events in the 1000 event sample, and should be read as
counting errors rather than strict percentages.}
\label{table1}
\end{table}

Whilst this efficiency is relatively high, it underestimates the actual
efficiency available to a real analysis.
The interference from delta electrons discussed in
Section~\ref{analysis:distances} means that an interest point related to the
primary vertex may be classified as closest to a secondary by the nearest
neighbour analysis.
However, a real analysis has no knowledge of the underlying physical features, 
and higher level algorithms will only require a rough localization of the 
primary vertex and other points. It is therefore more useful to ask ``is there
an interest point within a distance $d$ of the primary vertex?''. Whilst $d$ is somewhat 
arbitrary, $\sim5$mm was chosen as this provides good localization and matches 
the scale over which interference from delta electrons occurs. 
The number of events with an interest point within 5mm of the primary vertex was counted for each projection and divided by the total number of events, 
the results being shown in Table~\ref{table1}. This analysis was repeated to
determine the efficiency of finding an interest point within 5mm of one or both of the
muon and proton track ends, these results also being shown in
Table~\ref{table1}.

Both the primary vertex and muon/proton track ends are found with an efficiency
of $\sim90\%$ in both projections. Events with no
interest points close to the vertex or tracks ends were visually 
scanned to try and identify reasons for 
the failure. The major reason for failure to find the primary vertex
was a large opening angle between the muon and proton leading to local maxima in $R_N$ below threshold. A few events were also observed to have high energy
delta electrons close to the vertex that completely washed out the response from
the vertex.
With the muon/proton track ends, failure to find one or both was
caused by short proton tracks, projection of the tracks with very small
(projected) opening angle and again some delta electron contamination.
These are effects of the underlying physics though, and overall the primary
vertex and muon/proton track ends are found with excellent efficiency.

No matching of the two sets of 2D interest points to reconstruct
3D points is performed, but we can measure the efficiency for finding a feature
in one or both projections. Table~\ref{table2} shows combined efficiencies for 
finding an interest point $<5$mm from the primary vertex or muon/proton tracks 
ends in one or both projections. 
Compared with the $\sim90\%$ efficiency for finding the primary vertex/track 
ends in a single projection, the $\sim85\%$ combined efficiency is somewhat 
lower yet still performant.

\begin{table}
\begin{center}
\begin{tabular}{|l|c|}
\hline
Measure & Efficiency \\
\hline
Vertex Found In One or Both Projections & $(96.2\pm3.1)$\% \\
Vertex Found In Both Projections        & $(85.4\pm3.0)$\% \\
\hline
At Least One Track End In One or Both Projections & $(100.0\pm3.2)$\% \\
At Least One Track End In Both Projections & $(99.9\pm3.2)$\% \\
Both Track Ends Found In One or Both Projections & $(95.3\pm3.1)$\% \\
Both Track Ends Found In Both Projections & $(85.2\pm3.0)$\% \\
\hline
\end{tabular}
\end{center}
\caption{Efficiencies for detecting an interest point $<5$mm from the primary
vertex and muon/proton track ends in combined image projections. Uncertainties
were calculated as the Poisson error on the counted number of events in the 1000
event sample, and should be read as a counting error rather than strict
percentages.} 
\label{table2}
\end{table}

It should however be possible to develop techniques for reconstructing 3D
interest points from the two sets of 2D points. One could project the 
interest points as lines through the volume and find intersections and/or
close approches, but it may be better to perform a maximisation of the combined 
$R(x_1,x_2)$ response images. However, the optimum solution would be to
detect interest points directly from the full 3D voxel data as this contains 
the maximum information. Equation~\ref{structure:tensor} for the structure 
tensor extends naturally to three dimensions, but the extra degree of freedom 
means suitable response functions $R(x,y,z)$ are more sensitive to the exact 
structures being searched for.
Early work has identified several potential candidates for $R(x,y,z)$ to
identify vertex-like structures, and these will be reported in a future 
publication.

\subsection{Detection of Secondary Particles}
Whilst the primary vertex and track ends are the major key points that
higher level track finding algorithms would find useful, delta electron 
start/end points may also be needed. If the detector resolves these into 
separate tracks, certain track finding algorithms, especially those operating 
locally hit by hit, may need these interest points in order to 
identify branching points.

\begin{figure}[ht]
\begin{center}
\includegraphics[width=.75\textwidth]{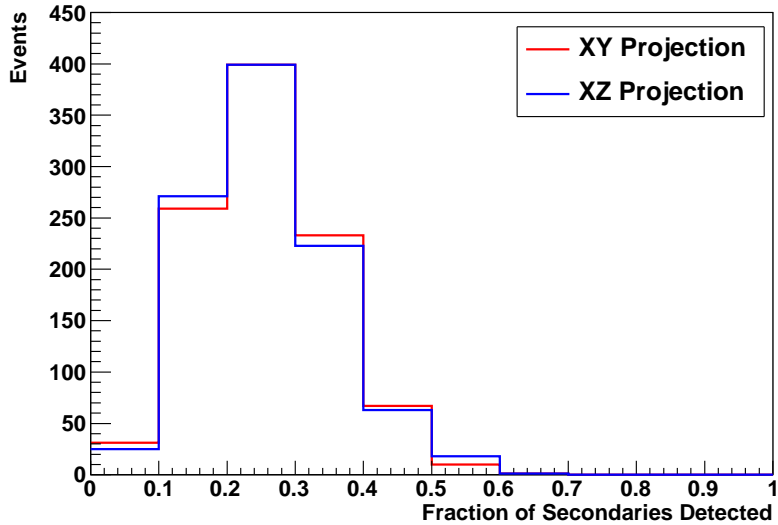}
\end{center}
\caption{Distribution of per-event fraction of found secondary start/end
points in the two image projections.}
\label{fig:secondary_eff}
\end{figure}

Delta electrons can appear as `T/Y' structures and are hence detectable as
interest points, as can be seen in Figure~\ref{event}. Their relatively low energy leads to a wide
variance in structure however, and hence their detectability as interest points
is strongly dependent on whether they clearly branch from their parent or simply
merge into it.

Figure~\ref{fig:secondary_eff} shows the distribution of the fraction of all
delta electrons in each event found as interest points. Clearly, delta electrons
can be detected, but only a relatively small fraction of the total are picked
out. This reflects the variance in delta electron structure rather than a 
failure of the interest point detection, and should not be taken as a direct
measure of the efficiency for finding delta electrons. That measurement
requires knowledge of the downstream reconstruction algorithm(s) as the
efficiency in this case would be quantified as the fraction of delta electrons
\textit{relevant to the downstream algorithm} that were detected. Not all delta
electrons will be relevant to downstream algorithms, e.g. if they merge into the
parent track. This study is left to later implementations of interest point
detection in a full reconstruction chain.

\section{Conclusions}
\label{conclusions}
This work has studied the application of a structure tensor based interest
point detection algorithm to find primary vertices and track end points in neutrino
charged current events recorded by a LAr-TPC. Interest point detection replaces the role
of a human operator in selecting keypoints for use by higher level
reconstruction, potentially providing a key part of a fully automated framework
for reconstruction.

A sample of charged current events was simulated in a Geant4
modelled LAr-TPC, with physical interest points and a voxel image of the 
energy deposition recorded. Two dimensional images were created from the voxel
image to match the reconstruction approach of ICARUS and ArgoNeut, with each
image analysed with the Noble response function to extract interest points. It
was found that the extracted interest points strongly correlate in position with
physical features, with 93\% of interest points being within 5mm of a physical
feature in both projections. Noise, gradient based effects, and interference between nearby physical
features cause a spread in the distribution of distances between interest points
and underlying physical features. Better localization may be possible by tuning
the derivative/averaging kernel widths used in calculation of the structure
tensor. Nevertheless, the localization achieved is likely to be sufficient for
most applications.

The efficiency with which interest points related to the primary vertex and
track ends were found was also measured. Initially, the criterium for finding
the primary vertex was taken as having an interest point whose nearest physical
feature was the primary vertex. This resulted in $\sim82\%$ of events finding the
primary vertex, independent of the projection. It was recognized that real applications would not need such a
strict criterium, since they simply require an interest point to be found close
to the primary vertex. Further analysis showed that $\sim91\%$ of events had an
interest point within 5mm of the primary vertex, independent of the projection.
The increase in efficiency arises from delta electrons close to the vertex 
interfering with the response, resulting in the nearest neighbour analysis 
classifying the interest point as a secondary, even though it is very close to
the 
vertex. An identical analysis showed that $\sim90\%$ of events found interest points
within 5mm of both the muon and proton track ends, independent of projection. 
Whilst no 3D reconstruction of interest points from the two sets of 2D points
was attempted, the efficiency analyses were combined to find the efficiencies
for finding the primary vertex and track ends in either or both projections for
the same event. This showed that the primary vertex or both track ends are found
in both projections in $\sim85\%$ of events. If real applications require better
localization of interest points related to the primary vertex and tracks ends,
this efficiency would decrease.

The application of interest point detection to identify delta electrons was
studied, with 10-60\% of secondaries detected as interest points in each
projection. Clearly, delta electrons can be detected as interest points, but it
is noted that this number does not represent an efficiency for detection as it
does not distinguish secondaries that will be relevant to reconstruction, e.g.
as a separate track, from those that simply merge into the main track. This
study is left to actual applications with a defined downstream algorithm.

Clearly, interest point detection provides an accurate and efficient method
for extracting physics-based points of potential use by other reconstruction
algorithms. Whilst this work has concentrated on their application to neutrino
interactions in a LAr-TPC, any detector with high granularity readout and
complex event topologies may find the technique useful. The initial
implementation presented has worked with two 2D projections of the underlying 
3D voxel data to give direct application to the reconstruction techniques
already implemented for LAr-TPCs. However, the structure tensor approach to
interest point detection should extend naturally to 3D, potentially giving
improvements in localization and efficiency from the extra information
available. Work has identified several suitable functions of the structure
tensor for identifying tracks and vertices in 3D, and will be reported in a
future publication.

\section{Acknowledgements}
The author is supported by STFC, and is grateful to Yorck Ramachers, Gary
Barker and Paul Harrison for useful discussions. The use of the SciPy package~\cite{scipy} in the
implementation of the interest point detection algorithms is acknowledged.



%% Bibliography
\bibliographystyle{JHEP}
\bibliography{hsp_bibliography.bib}

\end{document}